\documentclass[twocolumn,aps,floats,showpacs,superscriptaddress]{revtex4-2}
\usepackage[latin3]{inputenc}
\usepackage[makeroom]{cancel}
\usepackage{graphicx}
\usepackage{amsmath}
\usepackage{amsfonts}
\usepackage{amssymb}
\usepackage{color}
\usepackage[left=2cm,right=2cm,top=2cm,bottom=2cm]{geometry}
\usepackage{graphicx}
\usepackage{appendix}
\usepackage{ulem}
\DeclareFontFamily{OT1}{pzc}{}
\DeclareFontShape{OT1}{pzc}{m}{it}{ <-> s*[1.2] pzcmi7t }{}
\DeclareMathAlphabet{\mathpzc}{OT1}{pzc}{m}{it}

\begin{document}

\title{Adiabatic self-vibrations generated by inelastic Andreev tunneling in a movable Cooper-pair box}
\author{S. Park}
\email{sunghun.park@ibs.re.kr}
\affiliation{Center for Trapped Ion Quantum Science, Institute for Basic Science (IBS),
Daejeon 34126,
Republic of Korea}
\author{A. V. Parafilo}
\email{anton.parafilo@matfyz.cuni.cz}
\affiliation{Center for Theoretical Physics of Complex Systems,
Institute for Basic Science (IBS), Daejeon 34126,
Republic of Korea} 
\affiliation{Department of Condensed Matter Physics, Faculty of Mathematics and Physics,Charles University, Ke Karlovu 5, CZ-121 16 Prague, Czech Republic}
\author{L. Y. Gorelik}
\affiliation{Department of Physics, Chalmers University of Technology, SE-412 96 G{\" o}teborg, Sweden}
\author{R. I. Shekhter}
\affiliation{Department of Physics, University of Gothenburg, SE-412 96 G{\" o}teborg, Sweden}
\date{\today}

\begin{abstract}
Self-sustained oscillators produce stable periodic motion robust to dissipation. Such motion is usually achieved by work fed back into the oscillator, but its performance is often limited by frequency-dependent operation. Here we propose a scheme for self-sustained vibrations without external feedback. We consider a movable Cooper-pair box attached to the free end of a voltage-biased normal-metal pillar. The Cooper-pair box carries an Andreev current subject to an electric field applied perpendicular to the current. In the adiabatic limit, where the Cooper-pair box state follows its motion, vibrational instability occurs, pumped by inelastic Andreev tunneling. Nonlinearity of the Josephson coupling saturates the vibrational amplitude, resulting in two-dimensional self-vibrations. We discuss the advantage of this adiabatic scheme in comparison with feedback-induced self-oscillation.
\end{abstract}
\maketitle

\section{Introduction} 
Nanoelectromechanical systems (NEMS) integrate vibrating mechanical elements with electronic circuits on the nanoscale~\cite{Cleland03,Ekinci05,Bachtold22}, enabling applications from highly sensitive mass~\cite{Yang06} and force~\cite{Bachtold13} detection to elements of quantum information devices \cite{LaHaye09,Kim20}. 
However, mechanical vibrations are prone to dissipation and typically require AC driving with large external components, limiting scalability and nanoscale integration~\cite{Zettl10}. Consequently, there has been much interest in developing self-sustaining oscillators powered only by DC voltage, including electron shuttle and tunneling devices~\cite{Zant09,Blick10,Zettl10,Weig12,Baugh20}, as well as feedback-driven NEM resonators~\cite{Roukes08}.

Theoretical studies of the electron shuttle elucidate how coupling between mechanical motion and electric currents induces self-oscillation of a metal island between normal electrodes~\cite{Gorelik98,Jonson03}. At sufficiently high DC bias voltage, it exhibits shuttle instability with the oscillation amplitude growing until it reaches a limit cycle. 
Such instability can be viewed as a result of the effective negative friction force arising from the delay between the island charge and its position, as the charge does not fully follow the island motion due to retardation effect. This leads to a non-adiabatic correction, proportional to the island velocity, to the force, making the work over one oscillation period dependent on the oscillation frequency~\cite{Radic11,Radic13}. This dependence, which also arises in feedback-driven NEM resonators, may not supply sufficient work to overcome mechanical dissipation, limiting its use at low frequencies.

A movable Cooper-pair box (CPB) is a superconducting island whose mechanical motion modulates its Josephson coupling to a bulk superconductor~\cite{Gorelik01}. The coupling is position dependent, inducing a Josephson force depending on the CPB state, distinct from an electrostatic force. When the island is additionally coupled to a normal metal, Andreev tunneling produces a current through transforming two electrons into a Cooper pair on the island, thereby changing the state~\cite{Park25}. These interplay between the electronic and mechanical degrees of freedom has not yet been exploited. 

\begin{figure}[t!]
    \centering
    \includegraphics[width=0.9\linewidth]{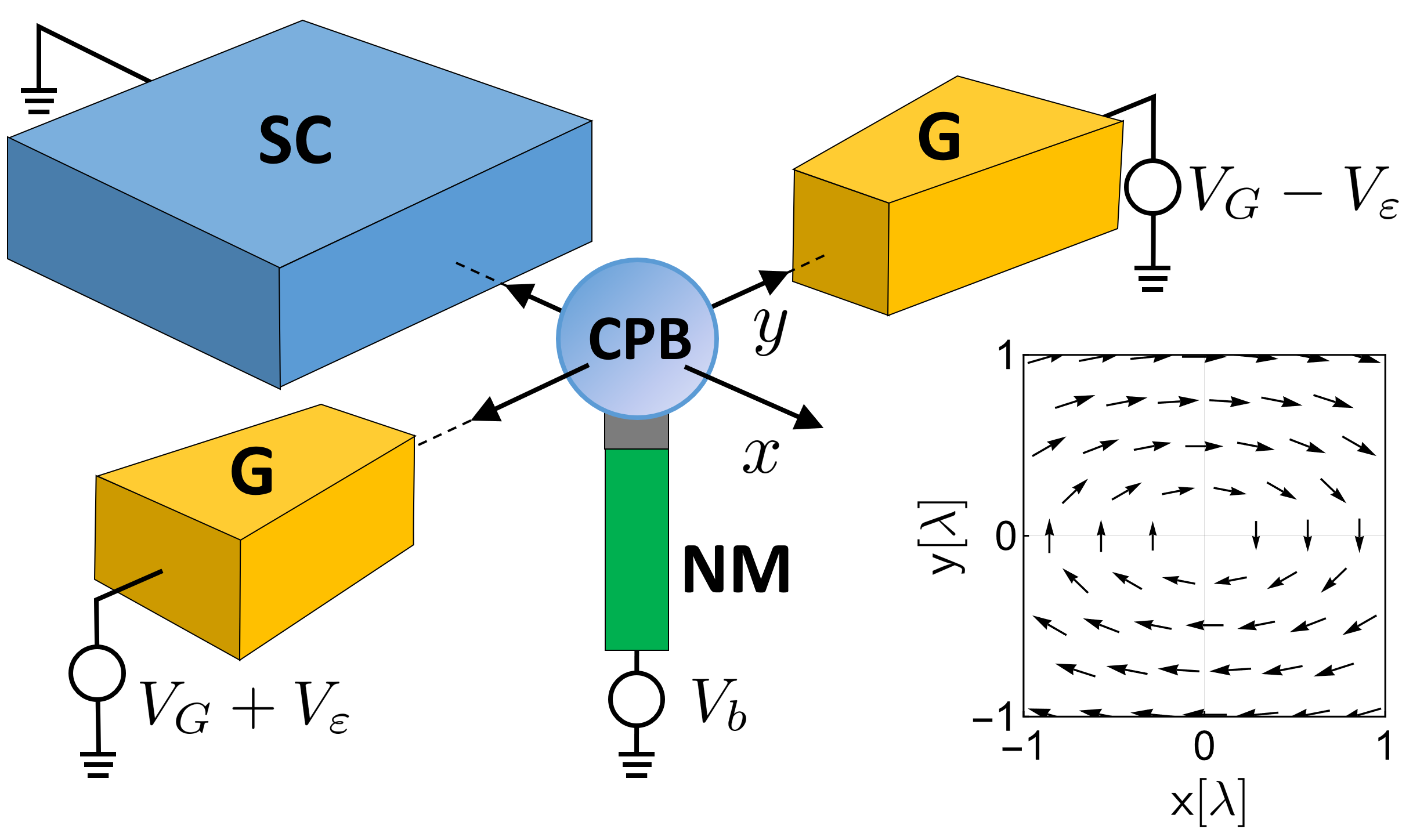}
    \caption{Schematics of the CPB device in the $xy$ plane. A movable CPB is mounted on the free end of a singly clamped normal-metal(NM) pillar biased at $V_b$, enabling in-plane mechanical oscillations accompanied by inelastic Andreev tunneling. It is tunnel-coupled to a superconducting (SC) electrode along $x$ and is influenced by two side gates (G) along $y$, biased at $V_G\pm V_{\varepsilon}$, where $V_G$ shifts the electrostatic potential of the CPB, while $V_{\varepsilon}$ generates the electric field $\mathcal{E}$. 
    Bottom right inset: schematic force field {\bf f}({\bf r}) in Eq.~\eqref{curlf2}. 
    }
    \label{fig1:setup}
\end{figure}

In this work, we show that self-vibrations of a movable CPB can occur at low frequencies. The CPB is attached to the free end of a voltage-biased normal-metal pillar, and carries a current through inelastic Andreev tunneling, in the presence of an electric field perpendicular to the current, see Fig.~\ref{fig1:setup}; a similar structure, without superconductivity, was implemented in Ref.~\cite{Kim12}. In the absence of tunneling, the CPB motion remains damped and decoupled from its electronic dynamics, which is governed by Josephson and electrostatic couplings. When tunneling is switched on, it adiabatically pumps work to the motion, giving rise to curl forces that drive the CPB into vibration instability. The vibrational growth is saturated by the nonlinearity of the Josephson coupling, resulting in self-sustained vibrations. This adiabatic mechanism without external feedback exhibits a frequency dependence different from that of feedback-based mechanisms, enabling more efficient development of instability.    

The paper is organized as follows. In Sec.~\ref{sec2}, we introduce the model Hamiltonian of the proposed nanoelectromechanical system and derive the equations for the reduced density matrix describing the dynamics of the movable CPB. We focus on the regime of small displacements, where the Josephson coupling can be represented in a linearized form, allowing analytical expressions for the force and the resulting instability condition to be obtained in terms of experimental parameters. Section~\ref{sec3} presents the resulting adiabatic self-sustained oscillation of the CPB. In Sec.~\ref{sec3a}, we discuss the role of noncommutativity between the Josephson and electrostatic couplings by calculating the curl forces in the vicinity of the static position and analyzing the resulting vibrational instability. In Sec.~\ref{sec3b}, we show the numerical results for the self-sustained oscillation of the CPB including the nonlinearity of the Josephson coupling. We discuss possible experimental signatures of the mechanical motion in current measurements in Sec.~\ref{sec3c}. Finally, Sec.~\ref{sec4} discusses experimental feasibility and summarizes our work.    

\section{Model Hamiltonian}\label{sec2}
The Hamiltonian of our setup depicted in Fig.~\ref{fig1:setup} is 
\begin{equation}
\hat{H} = \hat{H}_{\text{s}} +\hat{H}_{\text{n}}+\hat{H}_{\text{A}} +\hat{H}_{\text{m}}+\hat{H}_{\text{in}}. \label{htotal}
\end{equation}
The term $\hat{H}_{\text{s}}$ corresponds to the displacement-independent part of the CPB Hamiltonian, obtained by performing a Taylor expansion of the position-dependent Josephson coupling around the origin in the $xy$ plane. In the absence of mechanical displacement, the Hamiltonian takes the form of charge qubit for a superconducting island coupled to a bulk superconductor via the Josephson energy $E_J$~\cite{Wendin07, Devoret04}, in the regime where the Coulomb blockade is removed by the gate voltage $V_G=e/C$ with $C$ the island capacitance~\cite{Matveev93},   
\begin{equation}
\hat{H}_{\text{s}}= -E_{J} \hat{\sigma}_{1}, \label{hcpb}
\end{equation}
where $\vec{\sigma}=(\hat{\sigma}_{1},\hat{\sigma}_{2},\hat{\sigma}_{3})$ denotes the vector of Pauli matrices acting on the qubit subspace spanned by the ground state $|0\rangle=(0,1)^T$ and the charged state with a single Cooper pair $|1\rangle = (1,0)^{T}$. 
The island is attached to the normal-metal pillar with the Hamiltonian,
\begin{equation}
\hat{H}_{\text{n}}=\sum_{k,\kappa}(\varepsilon_k-e V_b) a_{k \kappa}^{\dagger} a_{k \kappa}, \label{hn} 
\end{equation}
where $a^{\dagger}_{k\kappa}$ creates an electron of energy $\varepsilon_k$ with momentum $k$ and spin $\kappa=\uparrow, \downarrow$ in the pillar. $V_b$ is the bias voltage. 
At the junction between the island and the pillar, electron exchange occurs via inelastic Andreev tunneling, which is modeled by   
\begin{equation}
\hat{H}_{\text{A}}=t_{\text{A}}\sum_{k,k'}\left(a^{\dagger}_{k\uparrow}a^{\dagger}_{k'\downarrow}\hat{\sigma}^{-}+a_{k'\downarrow}a_{k\uparrow}\hat{\sigma}^{+}\right),  \label{ha} 
\end{equation}
where $\hat{\sigma}^{\pm}=(\hat{\sigma}_{1}\pm i\hat{\sigma}_{2})/2$. This term takes into account the pair-electron process only under the assumption that the thermal broadening of the Fermi-Dirac distribution at temperature $T$ in the pillar is small compared to the SC gap, $k_B T \ll \Delta$, to suppress single-electron tunneling. 
The tunneling coefficient $t_{\text{A}}$ in Eq.~\eqref{ha} is proportional to $R_0/R_n$, where $R_n$ is the normal-state resistance of the junction and $R_0=h/e^2$ is the resistance quantum~\cite{Hekking93}.
We assume a constant $t_{\text{A}}$ during the mechanical motion. The mechanical degree of freedom of the combined CPB and the pillar, with frequency $\omega_0$, is described by  
\begin{equation}
\hat{H}_{\text{m}} = \frac{\hat{\bf{p}}^2}{2 m^*} +  \kappa\, \frac{\hat{\bf{r}}^2}{2}, \label{hm}
\end{equation}
where $\hat{\bf{r}}=\hat{x}{\bf{e}}_{x}+\hat{y}{\bf{e}}_y$ and $\hat{\bf{p}}={\hat{p}}_x{\bf{e}}_x + \hat{p}_y{\bf{e}}_y$ are position and momentum operators, respectively, and $\kappa=m^* \omega^2_0$ is the mechanical rigidity with the effective mass $m^{*}$. 
Note that $\kappa$ is controlled by the curvature of the mechanical potential at equilibrium, which depends on the elastic properties of the normal pillar and its geometry~\cite{LaHaye09, Zettl10}.
The term $\hat{H}_{\text{in}}$ reflects the coupling between the mechanical and electronic degrees of freedom of the CPB. 
To discuss the underlying mechanism, we focus on the small vibration in the vicinity of the origin where the coupling Hamiltonian can be linearized as
\begin{align}
\hat{H}_{\text{in}} = \frac{E_J}{\lambda} \, \hat{\bf{r}}\cdot\hat{\bf{F}}, \qquad
 \hat{\bf{F}}&=\hat{\sigma}_1  {\bf e}_x + \eta\,\hat{\sigma}_3{\bf e}_y,  \label{hin}
\end{align}
where $\eta=e\mathcal{E} \lambda/E_{J}$. This term characterizes the linearized distance dependence of the Josephson energy along the $x$ direction, with the decay length $\lambda$, and the change in the electrostatic energy in the $y$ direction under the electric field $\mathcal{E}$ from the side gates (see Fig.~\ref{fig1:setup}). 

We use a reduced density matrix approach to derive the dynamics of the CPB from the Liouville-von Neumann equation $\dot{\hat{\varrho}}_{\text{tot}}=-(i/\hbar) [\hat{H},\hat{\varrho}_{\text{tot}}]$ for the total density matrix $\hat{\varrho}_{\text{tot}}$, where the overdot denotes time derivative.
To this end, we employ the Born-Markov approximation, which is applicable in the parameter regime, 
\begin{equation}
  \Delta > e V_b \gg k_B T,\, \hbar\Gamma,\, E_J, \nonumber    
\end{equation}
where $\Gamma\equiv 2\pi t^2_{\text{A}} \nu^2|eV_b|/\hbar$, with $\nu$ the density of states in the pillar, is the tunneling rate. In this regime, any correlations within the pillar are quickly suppressed. Moreover, the coupling between the electronic and mechanical degrees of freedom characterized by $\epsilon$, as can be seen in Eq.~\eqref{eom1} below, is weak,  
\begin{equation}
  \epsilon \equiv  \frac{E_J}{ \kappa \lambda^2} \ll 1,  \label{em_coupling}
\end{equation}
This can be expressed as $\epsilon=(x_0/\lambda)^2 E_J/\hbar \omega_0$, where $x_0=\sqrt{\hbar/m^* \omega_0}$ is the zero-point fluctuation amplitude. For moderate values of $E_J/\hbar \omega_0$, such as those used in Fig.~\ref{fig2:selfoscillation}, Eq.~\eqref{em_coupling} implies that $x_0\ll \lambda$. Consequently, quantum fluctuations of the mechanical motion remain much smaller than the scale of the oscillation amplitude $\lambda$, allowing the mechanical motion to be treated semiclassically. Thus, quantum correlations between the electronic and mechanical degrees of freedom can be neglected.  
Then the total density matrix can be factorized as $\hat{\varrho}_{\text{tot}} = \hat{\varrho}_{\text{s}} \otimes\hat{\varrho}_{\text{m}} \otimes \hat{\varrho}^{\text{eq}}_{\text{n}}$, where $\hat{\varrho}_{\text{s}}$ and $\hat{\varrho}_{\text{m}}$ are the density matrices for the electronic states and mechanical motion of the CPB, respectively, and $\hat{\varrho}^{\text{eq}}_{\text{n}}$ is the equilibrium density matrix of the pillar.   
This yields a semiclassical treatment of the motion, while retaining quantum mechanical description of the CPB state, 
\begin{equation}
\ddot{{\bf r}}+\gamma_{\text{diss}}\dot{{\bf r}}+\omega_0^2{\bf r}={\bf f}({\bf r}), \,\,\,\,\,\,\,\,  
{\bf f}({\bf r})= -\epsilon\, \omega^2_0\,   \text{Tr}(\hat{\bf{F}}\hat{\varrho}_{\text{s}}), \label{eom1}
\end{equation}
where ${\bf r} = (1/\lambda)\text{Tr}(\hat{{\bf r}} \hat{\varrho}_{\text{m}})$ is the classical position normalized by $\lambda$, and $\gamma_{\text{diss}}$ is the mechanical dissipation rate. Here $\hat{\varrho}_{\text{s}}$ obeys the equation~\cite{Park25, Breuer02}, 
\begin{equation}
  \hbar \dot{\hat{\varrho}}_{\text{s}} =   i E_J [ \hat{\sigma}_1-{\bf r}\cdot\hat{{\bf F}} ,\hat{\varrho}_{\text{s}}] 
     - \hbar \Gamma \mathcal{L}[\hat{\sigma}^+] \hat{\varrho}_{\text{s}} , \label{Lindblad1}
\end{equation}
with $\mathcal{L}[\hat{\sigma}^+] \hat{\varrho}_{\text{s}} =  \{\hat{\sigma}^- \hat{\sigma}^+,\hat{\varrho}_{\text{s}} \} - 2 \hat{\sigma}^+ \hat{\varrho}_{\text{s}} \hat{\sigma}^-$ the Lindbladian operator, obtained by tracing out the pillar degrees of freedom, for $\text{sgn}(e V_b)>0$. The matrices $\hat{\sigma}^+$ and $\hat{\sigma}^-$ are exchanged when $\text{sgn}(e V_b)<0$. In what follows, we assume $\text{sgn}(e V_b)>0$ unless specified otherwise.   
The static solution is 
\begin{equation}
\begin{split}
{\bf r}&={\bf r}_{\text{st}}= -\epsilon \eta t^2_{n} {\bf e}_y + \mathcal{O}(\epsilon^2),  \\
  \hat{\varrho}_{\text{s}}&=\hat{\varrho}_{\text{st}}=\frac{1}{2}I+ \frac{1}{2} t_{n}\left(2 t_{J}\hat{\sigma}_2 + t_{n} \hat{\sigma}_3\right) +\mathcal{O}(\epsilon) ,  
\end{split}
\end{equation}
where $t_J$ and $t_n$ are dimensionless tunneling parameters associated with the superconductor and the pillar, respectively, 
\begin{equation}
  t_{J} =  \frac{E_J}{\sqrt{\hbar^2 \Gamma^2 + 2 E^2_J}}, \,\,\,\,\,
  t_{n} =  \frac{\hbar \Gamma}{\sqrt{\hbar^2 \Gamma^2 + 2 E^2_J}}. \nonumber 
\end{equation}
In the following, we consider the stability of the solution with respect to small deviations of $\delta{\bf r} \ll 1/\epsilon$ from ${\bf r}_{\text{st}}$.

\section{Adiabatic self-sustained oscillations}\label{sec3}
\subsection{Instability induced by curl forces}\label{sec3a}
To study the stability of the static solution, we solve Eqs.~\eqref{eom1} and \eqref{Lindblad1} in the adiabatic regime, $\omega_0 \ll \Gamma, E_J/\hbar$, where the vibration is much slower than the establishment of the steady Andreev current. In this regime, the CPB state follows its motion, which is therefore governed by the linearized dynamics around the static solution. We thus shift the origin of the $xy$-space to ${\bf r}_{\text{st}}$ and expand $\hat{\varrho}_{\text{s}}$ to linear order in ${\bf r}$: $\hat{\varrho}_{\text{s}} =\hat{\varrho}_{st} + {\bf r}\cdot\vec{\varrho}$, where $\vec{\varrho}=\hat{\varrho}_x{\bf e}_x +\hat{\varrho}_y{\bf e}_y$ (see Appendix~\ref{app1} for details). Substituting this into Eqs.~\eqref{eom1} and \eqref{Lindblad1} and imposing the stationary condition $\dot{\hat{\varrho}}_{\text{s}} =0$ yields ${\bf f}({\bf r}) = -\epsilon\omega^2_0 \text{Tr} [\hat{{\bf F}}({\bf r}\cdot\vec{\varrho})]$, a two-dimensional force field with nonzero curl, as 
\begin{align}
({\bf \nabla} \times {\bf f} ({\bf r}))\cdot {\bf e}_z\big|_{{\bf r}=0}  
&= -i \epsilon \omega^2_0 \, \text{Tr} \big[[\hat{F}_x , \hat{F}_y] \hat{R}\big] \label{curlf1}\\
&= -4 \eta \epsilon \omega^2_0 t^2_J (1+t^2_n), \label{curlf2}
\end{align} 
where $\hat{R}= t^2_J (1+t^2_n) \hat{\sigma}_2+ t_J t^3_n \hat{\sigma}_3$. Eq.~\eqref{curlf1}, valid for general forms of $\hat{F}_x$ and $\hat{F}_y$, highlights the role of their commutation relation: the nonzero curl force is attributed to noncommutativity and vanishes when they commute. 
Using $\hat{F}_x=\hat{\sigma}_1$ and $\hat{F}_y=\eta \hat{\sigma}_3$, as defined in Eq.~\eqref{hin}, we obtain ${\bf f} ({\bf r}) = 4 \epsilon \eta \omega^2_0 t^2_J (y {\bf e}_x - t^2_n x {\bf e}_y)$ and Eq.~\eqref{curlf2}, which shows that the sign of $\eta$, and thus of $\mathcal{E}$, determines the direction of the rotational force, clockwise for $\eta>0$ and counterclockwise for $\eta<0$ (see the inset of Fig.~\ref{fig1:setup}).    
Note that the curl force cannot be derived from a scalar potential~\cite{Berry15} and reflects the quantum nature of the CPB state arising from the position-dependent Josephson coupling, which has no classical analogue. 

The inelastic Andreev tunneling, characterized by $\Gamma$, is essential for generating the curl force. 
From Eqs.~\eqref{eom1} and \eqref{Lindblad1}, we find that the force ${\bf f} ({\bf r})$ decouples the mechanical motion along the $x$ and $y$ directions, i.e., ${\bf f} ({\bf r}) \sim f_x(x) {\bf e}_x + f_y(y) {\bf e}_y$, when $\Gamma = 0$ and $\dot{\hat{\varrho}}_{\text{s}} =0$.
Consequently, the dynamics reduces to one-dimensional damped motion in each coordinate, with zero curl force. 

\begin{figure}
    \centering
    \includegraphics[width=0.9\linewidth]{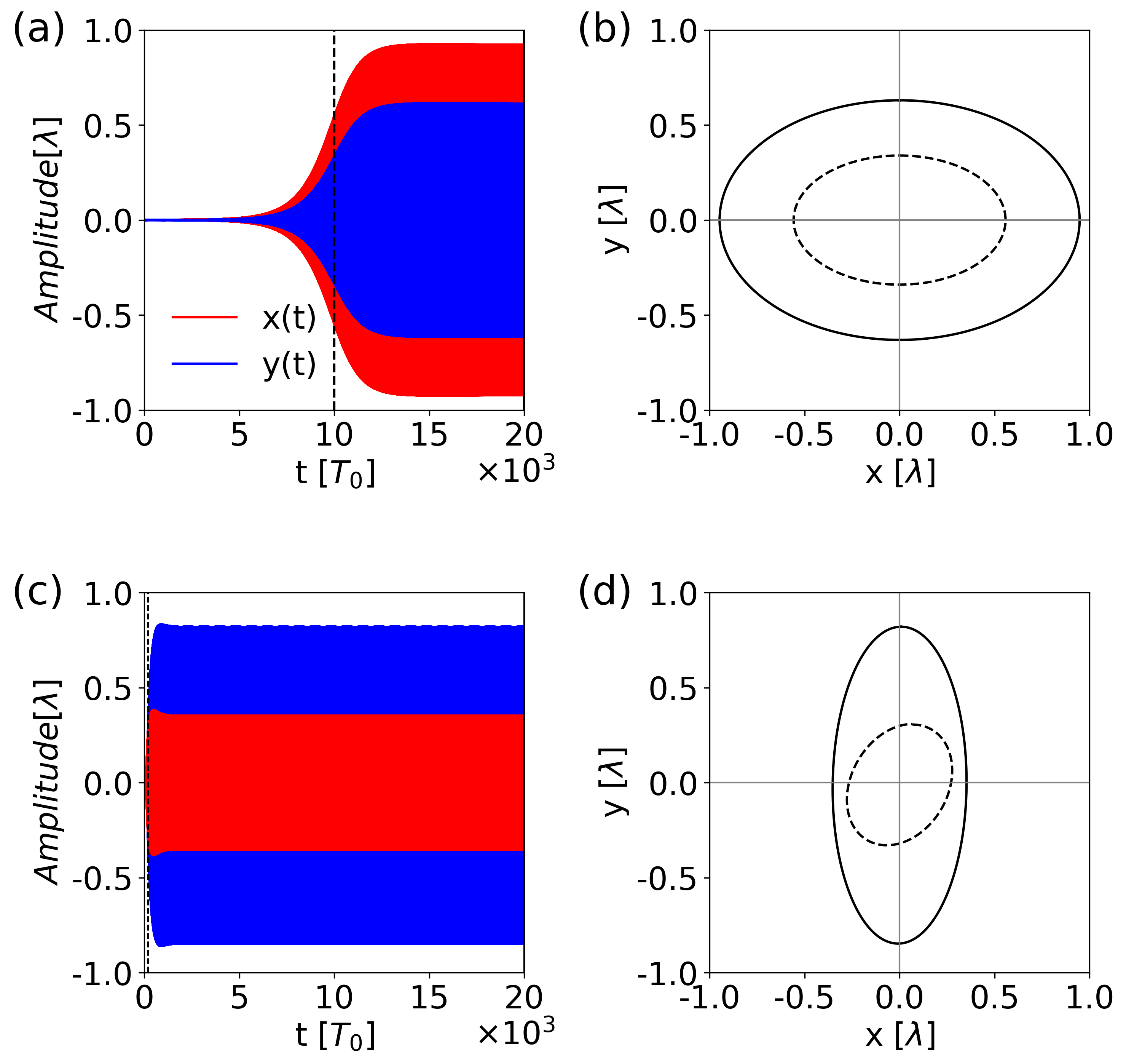}
    \caption{Self-oscillation of the CPB for the initial condition $(x,y,\dot{x},\dot{y})=0$ at $t=0$, with $E_J=\hbar\Gamma=150 \,\hbar\omega_0$, $\epsilon=1.5\times10^{-3}$, $Q$-factor $Q=\omega_0/\gamma_{\text{diss}}=10^{3}$, and $\omega_0/2\pi=100 \,\text{MHz}$. 
    (a) For $\eta=1.1$, the amplitudes $x(t)$ (red) and $y(t)$ (blue), in units of $\lambda$, are plotted versus time $t$ in units of the oscillation period $T_0=2\pi/\omega_0$. (b) Single-cycle motion in the $xy$ plane at $t=10^4 \,T_0$ (dashed) and $2\times10^4 \,T_0$ (solid) from the oscillations in (a). 
    (c) Same as (a), but for $\eta=10$. (d) Same as (b), but from the oscillations in (c) at $t= 2\times10^2 \,T_0$ (dashed) and $2\times 10^4 \,T_0$ (solid).}
    \label{fig2:selfoscillation}
\end{figure}
We consider the work done by the force. 
In the adiabatic regime, the eigenvector of the characteristic equation for the frequency derived from Eq.~\eqref{eom1} sets a closed trajectory around the origin, parameterized as~\cite{Beletsky95}
\begin{equation}
  {\bf r}(t)= A(t)\, [\sin(\omega_0 t) \,{\bf e}_x+ \text{sgn}(\eta)\, t_n\,\cos(\omega_0 t) \,{\bf e}_y],  
\end{equation}
with slowly varying amplitude $\dot{A}(t)\ll\omega_0 A$. The work per cycle is evaluated by integrating along the trajectory, or equivalently by integrating Eq.~\eqref{curlf2} over the area enclosed by the trajectory. Including dissipation, we obtain the total work  $W=W_{\text{curl}}-W_{\text{diss}}$, where   
\begin{equation}
  W_{\text{curl}} = 4 \epsilon \omega^2_0 \eta \pi A^2(t)  t^2_J t_n (1+ t^2_n) \label{W_curl} 
\end{equation}
arises from the force ${\bf f}$ and $W_{\text{diss}}= \omega_0 \gamma_{\text{diss}} \pi A^2(t) (1+ t^2_n)$ from the dissipation.
If the work is positive, $W>0$, a vibrational instability develops and its amplitude grows exponentially. 
The work $W_{\text{curl}}$ increases linearly with the field $\mathcal{E}$, through the dependence on $\eta$, exhibits a maximum at $E_J/\hbar \Gamma \approx 0.8$, and vanishes as $E_J/\hbar \Gamma \rightarrow 0$ or $\infty$. The criterion can be expressed using the angular momentum $L=x\dot{y}-y\dot{x}$,  
\begin{equation}
\dot{L}=(\gamma_{\text{curl}}-\gamma_{\text{diss}}) L, \,\,\, \gamma_{\text{curl}} = 4 \epsilon \eta \omega_0 t^2_J t_n,
\end{equation} 
which indicates that $\gamma_{\text{curl}}-\gamma_{\text{diss}}$ is the pumping rate of the angular momentum.

We discuss the frequency dependence of the work. 
Since $\epsilon$ is inversely proportional to $\omega^2_0$, as defined in Eq.~\eqref{em_coupling}, $W_{\text{curl}} \propto \epsilon \omega^2_0$ in Eq.~\eqref{W_curl} is independent of $\omega_0$. As a result, $W_{\text{diss}}/W_{\text{curl}} \propto \omega_0$, implying that the dissipative work becomes negligible compared to the work done by the curl force in low-frequency limit.   
This dependence contrasts with the case in which feedback induces an instability through a retarded force ${\bf f}(t) = f_0 {\bf r}(t-\tau)$ with a delay time $\tau$, which leads to a single-cycle work $W_{\text{ret}}$ proportional to $\omega_0$ and to $W_{\text{diss}}/W_{\text{ret}}$ being independent of $\omega_0$. Therefore, our mechanism enables a more efficient occurrence of the instability.

\subsection{Self-oscillating motion of the Cooper-pair box}\label{sec3b}
We now take into account the nonlinearity in the Josephson coupling. 
We replace the linear dependence $-E_J (1-x)$ in Eq.~\eqref{htotal} with $-E_J e^{-x}$, and solve the nonlinear Eqs.~\eqref{eom1} and \eqref{Lindblad1} numerically, see Fig.~\ref{fig2:selfoscillation}. 
As shown in Eq.~\eqref{W_curl}, in the linear regime applicable near the origin, $x\ll1$, $W_{\text{curl}}$ scales as $A^2$. 
However, as the vibration amplitude grows and approaches $x\sim 1$, the nonlinear effects become important, and cause the work to grow nonlinearly with $A^2$ until it is balanced by the dissipative work, resulting in self-sustained vibrations.    

The results in Fig.~\ref{fig2:selfoscillation} show that the limit-cycle trajectory in the $xy$ plane is controlled by $\eta$. 
For small $\eta$, the motion is elongated along the $x$ direction (Fig.~\ref{fig2:selfoscillation} (b)), whereas an increase in $\eta$, and thus in the electric field, stretches it along the $y$ direction (Fig.~\ref{fig2:selfoscillation} (d)). In addition, the timescale required to reach the saturation regime decreases with increasing $\eta$. 

Along the limit cycle, a periodic exchange between the Josephson and electrostatic energies occurs. When the CPB passes through $y=0$, the Josephson coupling reaches its maximum, while the electrostatic energy $|e\mathcal{E}y|$ is minimized. As it deviates from $y=0$, the Coulomb blockade tends to be restored, suppressing the Josephson coupling. We will see below that this variation of the Josephson energy along the trajectory leads to more pronounced peaks in currents at larger $\eta$. 
\begin{figure}
    \centering
    \includegraphics[width=0.95\linewidth]{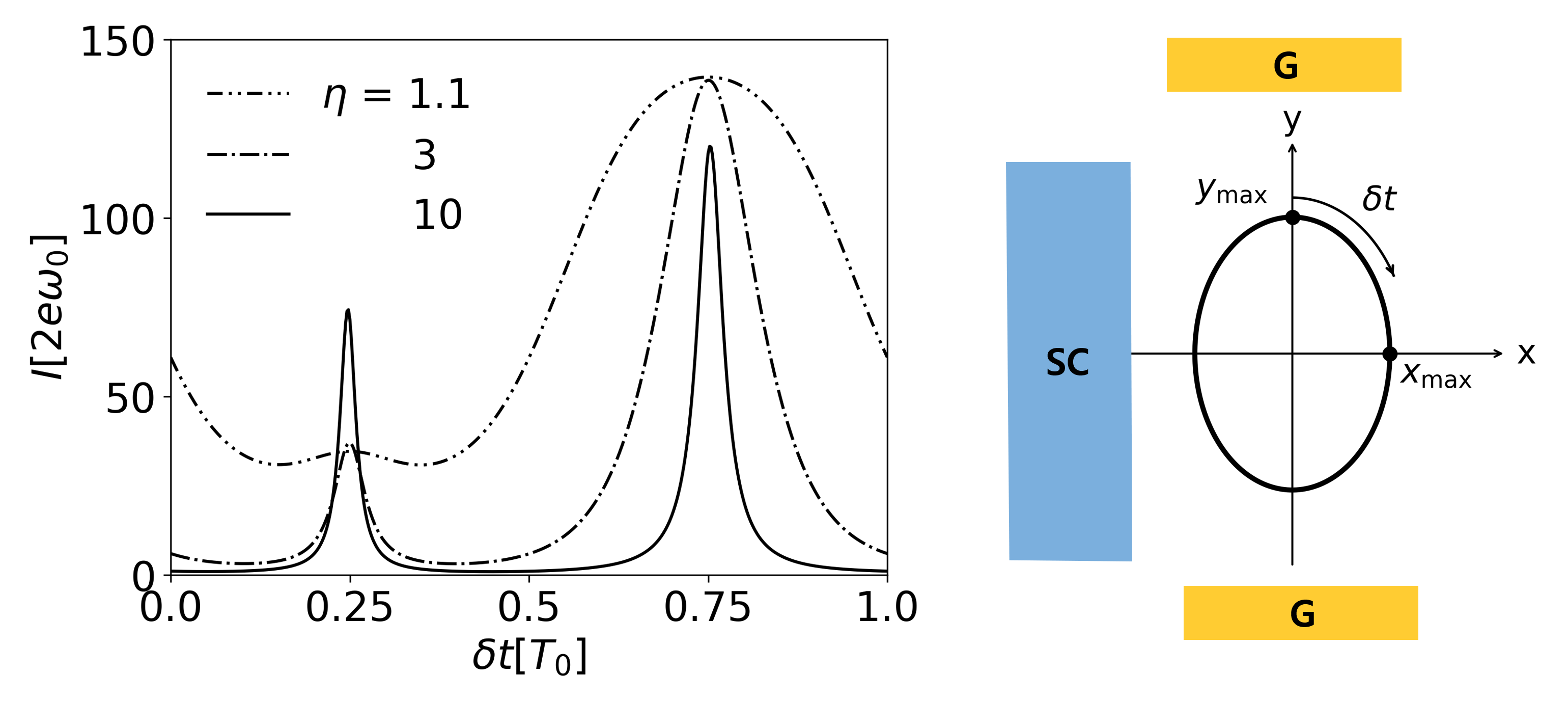}
    \caption{Left panel: Position-dependent adiabatic currents flowing from NM to SC through the CPB over one oscillation period in the self-oscillation limit for different values of $\eta$. The other parameters are the same as in Fig.~\ref{fig2:selfoscillation}. For clarity, the currents are plotted as a function of $\delta t$, which measures the elapsed time along the CPB trajectory from $(x,y)=(0, y_{\text{max}})$, as illustrated in the right panel. As $\eta$ increases, the current peaks at $\delta t = 0.25 \,T_0$ and $0.75 \,T_0$, at which the CPB positions $(x_{\text{max}},0)$ and $(-x_{\text{max}},0)$, respectively, become more pronounced, manifesting an electrical visualization of the mechanical motion.}
    \label{fig3:current}
\end{figure}

\subsection{Electric visualization of mechanical motion}\label{sec3c}
We calculate the time-dependent Andreev current $I= -e \langle\dot{\hat{\sigma}}_3 \rangle$ over one limit cycle. Fig.~\ref{fig3:current} shows the current, which is modulated due to the time-dependent Josephson coupling, for different values of $\eta$. Two characteristic maxima correspond the moments when the CPB reaches $y=0$, where the Coulomb blockade is lifted, while the current is suppressed elsewhere. The peaks become more pronounced as $\eta$ increases. Hence, these low-frequency current oscillations provide a direct means of visualizing the mechanical vibration. 

The $\eta$ dependence can be understood from the form of the adiabatic current, given by 
\begin{equation}
  I(x,y) =\displaystyle \frac{2e}{\hbar}\frac{2 \hbar \Gamma E^2_J e^{-2x} }{\hbar^2\Gamma^2+2 E^2_J e^{-2x} + 4 E^2_J \eta^2 y^2 }. \label{I_adiabatic} 
\end{equation} 
This explains that, due to the term $4 E_J^2\eta^2 y^2$, the current exhibits its maximal at $y=0$, corresponding to $\delta t/T_0 = 0.25$ and $0.75$ in Fig.~\ref{fig3:current}, where $T_0$ is the oscillation period. As $\eta$ increases, the distinction between the peak and the suppressed regions becomes clearer, sharpening the peaks.  
The relative height of the peaks can be expressed in terms of the half-amplitudes $x_{\text{max}}$ and $y_{\text{max}}$ along each axis, as shown in the right panel of Fig.~\ref{fig3:current}. In particular, for the case $E_J=\hbar \Gamma$, we obtain  
\begin{equation}
 \frac{I_1}{I_2}=\frac{e^{-2 x_{\text{max}}}+2}{e^{2 x_{\text{max}}}+2}, \,\, \frac{I_{\text{0}}}{\sqrt{I_1 I_2}}=\frac{\sqrt{5+4 \cosh{(2 x_{\text{max}})}}}{3+4 \eta^2 y^2_{\text{max}}}, \nonumber
\end{equation}
where $I_1$ and $I_2$ denote the current at $\delta t/T_0 = 0.25$ and $0.75$, respectively, and $I_{\text{0}}$ at $\delta t/T_0 = 0$. 
These relations reflect the current oscillation at the mechanical frequency and can be used to estimate parameters associated with the mechanical vibrations.  

\section{Discussion}\label{sec4}
The experimental realization of the proposed setup requires two key ingredients. The first is coherent Josephson coupling across a vacuum gap between a superconducting electrode and a movable superconducting island. Such coupling has been demonstrated in superconducting STM experiments~\cite{Padniuk26}. The second is the incorporation of the superconducting island into a mechanically compliant structure capable of sustaining nanometer-scale motion. Mechanical shuttle devices with oscillation frequencies of several hundred MHz and vibration amplitudes on the nanometer scale have already been demonstrated~\cite{Kim12}. Since these two essential ingredients have been realized separately, the proposed setup appears experimentally feasible with existing nanomechanical and superconducting technologies.

To estimate the experimental parameters, we take $\omega_0/2\pi=10^{8} \,\text{Hz}$ of $ m^*=10^{-18} \, \text{kg}$ CPB island, which has a zero-point fluctuation amplitude $\sqrt{\hbar/m^* \omega_0} \approx0.3 \,\text{pm}$, much smaller than the order of $\lambda=0.1\, \text{nm}$~\cite{Xu22}. 
Taking the rates $\Gamma, E_J/\hbar \approx 10^{11} \,\text{s}^{-1} \gg \omega_0$ to be in the adiabatic regime, $\eta$ needs to exceed $10^3 \,Q^{-1}$ and $T\ll 1 \, \text{K}$ for the instability~\cite{LaHaye09}. For the proposed parameters, the charge transferred per cycle $q$ is sufficiently large compared to charge fluctuations, e.g., $q\approx 70\times 2e$ for the current at $\eta=10$ in Fig.~\ref{fig3:current}.  

We predict a feedback-free nanomechanical instability, demonstrating that the non-commutativity between Coulomb and Josephson couplings induces adiabatic rotational motion of a CPB, with the instability supplied by inelastic Andreev tunneling. The predicted vibration and its parameter regime are accessible using standard Andreev-current measurement techniques. The instability remains robust in the presence of experimentally relevant gate-voltage fluctuations (see Appendix~\ref{app2}). This instability is expected to evolve into a self-sustained vibrations whose trajectory can be electrostatically controlled. The adiabatic nature enables a low-frequency self-sustained oscillator that can be integrated with superconducting circuits for hybrid superconducting NEMS. 

The predicted mechanism results in self-sustained vibrations even in a low-frequency regime owing to the frequency-independent work supplied by inelastic Andreev tunneling, eliminating the need for external feedback circuitry. Combined with the ability of the movable CPB to carry superconducting quantum states, the resulting motion may provide a platform for self-driven transport of quantum information between spatially separated superconducting elements. 
Thus, unlike a self-oscillating system operating in the nonadiabatic regime, where mechanical velocity can induce transitions between CPB states, the present mechanism enables spatial motion of the CPB while maintaining its evolution adiabatic. This combination of self-sustained motion and adiabatic evolution may provide a basis for mechanically a shuttling a qubit state between spatially separated superconducting elements while suppressing unwanted state transitions.
Generalization to multiple CPBs or the quantum regime of mechanical motion could enable mechanically mediated coherent interactions and entanglement between the islands.

\section*{Acknowledgment}

We gratefully acknowledge useful discussions with B. Altshuler, C. Kim, H. C. Park. S. P. acknowledges the
support from the Institute for Basic Science (IBS) in the Republic of Korea through the Project IBS-R024-Y4.
L. Y. G. and R. I. S. acknowledge the hospitality of the PCS at IBS, Republic of Korea, where part of this work
was supported by IBS Funding No. IBS-R024-D1. This work has been supported by QUANTERA ENTUNE, the
Swedish Research Council (VR 2025-08488), the Ministry of Education, Youth and Sports of the Czech Republic (Project ID: 9C26004), and co-
funded by the European Commission (Grant Agreement
No. 101212998). This research was also supported by the Czech Science Foundation through the Junior Star grant
25-17250M.

\appendix

\section{Derivation of Eq.~\eqref{curlf1}}\label{app1}
We assume $\hat{{\bf F}}=\hat{F}_{x} {\bf e}_x +\hat{F}_{y} {\bf e}_y$ and $\vec{\varrho}=\hat{\varrho}_x {\bf e}_x +\hat{\varrho}_y {\bf e}_y$, and parameterize $\hat{F}_{i=x,y}$ and $\hat{\varrho}_{i=x,y}$ in the Pauli basis as
\begin{equation}
\hat{F}_{i}=\sum_{j=1}^{3} F_{ij}\hat{\sigma}_{j},
\qquad
\hat{\varrho}_{i}=\sum_{j=1}^{3} \varrho_{ij}\hat{\sigma}_{j}. \nonumber
\end{equation}
Substituting these expressions into Eq.~\eqref{Lindblad1} and equating the coefficients of the Pauli matrices, we obtain
\begin{align}
\varrho_{i1} &= t_{n}t_{J}F_{i2}-2t_{J}^{2}F_{i3}, \nonumber\\
\varrho_{i2} &= t_{n}t_{J}(1-2t_{n}^{2})F_{i1},\\
\varrho_{i3} &= 2t_{n}^{2}t_{J}^{2}F_{i1}. \nonumber
\end{align}
Using the identity
\begin{align}
{\bf \nabla}\times \mathrm{Tr}\!\left[\hat{\mathbf F} ({\bf r}\cdot\vec{\varrho})\right]
&=-\mathrm{Tr} \left( \hat{\mathbf F}\times\vec{\varrho} \right) \nonumber\\
&= -\mathrm{Tr} \left( \hat{F}_{x}\hat{\varrho}_{y}- \hat{F}_{y}\hat{\varrho}_{x} \right){\bf e}_z, \nonumber
\end{align}
and substituting Eqs.~(A1) into the above expression, one finds 
\begin{equation}
-\mathrm{Tr} \left( \hat{F}_{x}\hat{\varrho}_{y} - \hat{F}_{y}\hat{\varrho}_{x} \right)=
i\,\mathrm{Tr} \!\left( [\hat{F}_{x},\hat{F}_{y}] \hat{R} \right),\label{curlidentity}
\end{equation}
where $\hat{R}= t^2_J (1+t^2_n) \hat{\sigma}_2+ t_J t^3_n \hat{\sigma}_3$. 
Equation~\eqref{curlidentity} immediately yields Eq.~\eqref{curlf1}.

\section{Effect of gate voltage fluctuations on the instability}\label{app2}

To account for fluctuations of the gate voltage $V_G$, we include dephasing and relaxation terms associated with gate voltage noise to the Lindbladian~\cite{Gambetta06}, 
\begin{equation}
 \Gamma \mathcal{L}[\hat{\sigma}^+] \hat{\varrho}_{\text{s}} \rightarrow 
 \Gamma \mathcal{L}[\hat{\sigma}^+] \hat{\varrho}_{\text{s}}+
 \gamma_1 \mathcal{L}[\hat{\sigma}^-] \hat{\varrho}_{\text{s}}+
 \frac{\gamma_{\varphi}}{2} \mathcal{L}[\hat{\sigma}_3] \hat{\varrho}_{\text{s}}, \label{app_b:lindblad}
\end{equation}
where $\gamma_1$ and $\gamma_{\varphi}$ denote the relaxation and dephasing rates of the CPB state, respectively. 
Here we assume sgn($eV_b$) $>$ 0: otherwise, $\gamma_1$ can be simply absorbed into the damping rate $\Gamma$. The gate voltage noise leads to the replacement of $t_J$ and $t_n$ by $\tilde{t}_J$ and $\tilde{t}_n$ in the expression for the force ${\bf f} ({\bf r})$ acting on the CPB, 
\begin{equation}
  {\bf f} ({\bf r}) = 4 \epsilon \eta \omega^2_0 \tilde{t}^2_J (y {\bf e}_x - \tilde{t}^2_n x {\bf e}_y),   
\end{equation}
where 
\begin{eqnarray}
    \tilde{t}_{J} =  \frac{E_J}{\sqrt{\hbar^2 \Gamma_a \Gamma_b + 2 E^2_J}}\sqrt{\frac{\Gamma_c}{\Gamma_a}}, \,\,\,\,\,\,\,
  \tilde{t}_{n} =  \frac{\hbar \Gamma_a}{\sqrt{\hbar^2 \Gamma_a \Gamma_b + 2 E^2_J}}, \nonumber 
\end{eqnarray}
and $\Gamma_a=\Gamma+\gamma_1+2 \gamma_{\varphi}$, $\Gamma_b=\Gamma+\gamma_1$, $\Gamma_c=\Gamma-\gamma_1$.

We estimate $\gamma_1$ using parameters of our model. At low temperature $k_B T \ll E_J$ and $\Gamma=0$, the relaxation rate of a CPB coupled to an Ohmic environment with resistance $R$ is given by \cite{Schoelkopf03} 
\begin{equation}
\frac{E_J}{\hbar \gamma_1}=\left(\frac{C_{\Sigma}}{C_g}\right)^2 \frac{R_K}{4\pi R},
\end{equation}
where $R_K=h/e^2\approx25.8 \,\rm k\Omega$ is the resistance quantum, and the capacitance ratio is $C_g/C_{\Sigma}=10^{-2}$. 
Taking realistic values of $R=377 \,\Omega$ \cite{Jack16} and $R=67.32 \,\Omega$ \cite{Padniuk26} give 
\begin{equation}
\frac{E_J}{\hbar\gamma_1}\approx 5.4\times10^4-3\times10^5.
\end{equation}
For $E_J/h=15\,\mathrm{GHz}$, this corresponds to $\gamma_1/(2\pi)\approx 0.05 \,\mathrm{MHz} -0.27\,\mathrm{MHz}$. The dephasing rate is $\gamma_{\varphi}/(2\pi)\approx 1 \,\mathrm{MHz}$ for typical CPB parameters \cite{Koch07}. 
These estimates show that the relaxation and dephasing induced by the environment are much smaller than the coherent Josephson tunneling characterized by $E_J/\hbar$, supporting the treatment of the environment as a perturbation and the description by additional Lindblad terms in Eq.~\eqref{app_b:lindblad}. 
Note that the robustness of the coherent Josephson coupling against the electromagnetic environmental originates from the removal of the Coulomb blockade around the charge degeneracy point of the CPB. This situation is distinct from incoherent Cooper-pair tunneling through a biased Josephson junction in the charging energy dominated regime, where the electromagnetic environmental effects are described by $P(E)$ theory~\cite{Ingold92, Jack16, Ko21}.

The damping rate $\Gamma$ arising from tunneling between the island and the normal-metal pillar is the order of $\Gamma\sim 10^{11}\, \rm{s}^{-1}$, which is much larger than typical relaxation and dephasing rates arising from charge noise reported for CPBs, which are, as discussed above, of order $10^5-10^6 \,\rm{s}^{-1}$. Therefore, in the parameter regime considered in our setup, the voltage induced relaxation and dephasing constitute only perturbation corrections to the dominant tunneling-induced dynamics already included in the model.

To quantify their effect, we evaluate the corrections to the work generated by the curl force $W_{\rm{curl}}=W^{(0)}_{\rm{curl}}+\delta W_{\rm{curl}}$ and the mechanical dissipation $W_{\rm{diss}}=W^{(0)}_{\rm{diss}}+\delta W_{\rm{diss}}$ within a perturbation treatment, where $W^{(0)}_{\rm{curl}}$ and $W^{(0)}_{\rm{diss}}$ denote the values in the absence of the gate voltage noise, while $\delta W_{\rm{curl}}$ and $\delta W_{\rm{diss}}$ are the corresponding corrections,
\begin{eqnarray}
 \delta W_{\rm{curl}}&=&  4 \epsilon \omega^2_0 \eta \pi A^2(t) \Big[ -4 t^2_J t_n (t^4_J+2 t^2_J t^2_n+2 t^4_n) \frac{\gamma_1}{\Gamma} \nonumber\\
 &&- 2 t^2_J t^3_n (2 t^2_n-t^2_J) \frac{\gamma_{\varphi}}{\Gamma}  \Big], \nonumber\\
 \delta W_{\rm{diss}} &=& \omega_0 \gamma_{\text{diss}} \pi A^2(t) \left[4 t^2_J t^2_n \frac{\gamma_1}{\Gamma}+2 t^2_n (1+2 t^2_J)\frac{\gamma_{\varphi}}{\Gamma} \right].\nonumber
\end{eqnarray}
The leading-order contributions show that the additional dephasing and relaxation terms suppress $W_{\rm{curl}}$ while increasing $W_{\rm{diss}}$. Nevertheless, these corrections remain sufficiently small and do not qualitatively alter the instability discussed in Sec.~\ref{sec3}.


\end{document}